\documentstyle[12pt,epsf,psfig]{article} 
\begin{document}
\title{  \bf Constraints of  the 4th
generation from $g_\mu -2$}
\author{Wu-Jun Huo\,\,and \,\,Tai-Fu Feng\\{\sl CCAST (World Lab.),
P.O. Box 8730, Beijing $100080$}\\{\sl  and} \\ {\sl
Institute of High Energy Physics, Academia Sinica, P.O. Box $918(4)$},\\{\sl
 Beijing $100039$, P.R.  China}}

\date{}
\maketitle

\begin{abstract}
We investigate the newly observed muon $g_\mu -2$ anomaly in the
framework of a sequential fourth generation model with a heavy
fourth neutrino, $\nu'$. We find that $g_\mu -2$ can exclude most
values of $m_{\nu'}$  and put a very stringent constraint on the
existence of the fourth generation. We also obtain bounds of a
$4\times 4$ leptonic mixing  matrix elements, $V_{2\nu'}$ if it
really exists.
\end{abstract}
\newpage

At present, Standard model (SM) has to face  the experimental
difficulties which are all relate to leptons. It seems to indicate
the presence of new physics just round the corner will be in the
leptonic part. Firstly, there are  convincing evidences that
neutrinos are massive and oscillate in flavor \cite{neutrino}.
Secondly, the recent measurement of the muon anomalous magnetic
moment by the experiment E821 \cite{e821} at Brookhaven National
Laboratory disagrees with the SM expectations at more than
2.6$\sigma$  level.

Defined as $a_\mu \equiv (g_\mu -2)/2$, the recent measurement of $a_\mu$ is
\begin{eqnarray}
a^{\rm exp}_\mu =(11\,659\,202\,\pm14\pm6)\times 10^{-10},
\end{eqnarray}
where the SM prediction is
\begin{eqnarray}
a^{\rm SM}_\mu =(11\,659\,159.7\,\pm6.7)\times 10^{-10},
\end{eqnarray}
Thus, one finds
\begin{eqnarray}
 a^{\rm exp}_\mu -a^{\rm
SM}_\mu =(42.6\pm16.5)\times 10^{-10}.
\end{eqnarray}
 which gives the old $2.6\sigma$ deviation. But the revised
difference between experiment and SM is
 \begin{eqnarray}
\delta a^{\rm SM}_\mu \equiv a^{\rm exp}_\mu -a^{\rm
SM}_\mu =26(16)\times 10^{-10}
\end{eqnarray}
which is now only a $1.6\sigma$ deviation \cite{new}. Although the
deviation drops from $2.6\sigma$ to $1.6\sigma$, it might seem
that the absolute magnitude of the deviation may be a hint of new
physics. There have been a lot of scenarios of new physics
proposed to interpret the non-vanishing and positive value of
$\delta a_\mu$. In this note, we consider the sequential fourth
generation standard model (SM4) to investigate its contributions
to $a_\mu$.

   From the point of phenomenology,
 there is a realistic question what are numbers of the fermions
generation or weather there are other additional quarks or
leptons. The present experiments  tell us there are only three
generation fermions with $light$ neutrinos which mass are less
smaller than $M_Z /2$\cite{Mark}. But the experiments don't
exclude the existence of other additional generation, such as the
fourth generation, with a $heavy$ neutrino, i.e. $m_{\nu_4} \geq
45.5 {\rm GeV}$\cite{Berez}. Many refs. have studied  the SM4
\cite{McKay}, which is added   an up-like quark $t^{'}$, a
down-like quark $b^{'}$, a lepton $\tau^{'}$,    and a heavy
neutrino $\nu^{'}$ in the SM.  The properties of these new
fermions
 are all the same as their corresponding counterparts
  of other three generations except their masses and CKM mixing, see Tab.1,
\begin{table}[htb]
\begin{center}
\begin{tabular}{|c||c|c|c|c|c|c|c|c|}
\hline
& up-like quark & down-like quark & charged lepton &neutral lepton \\
\hline
\hline
& $u$ & $d$& $e$ & $\nu_{e}$ \\
SM fermions& $c$&$s$&$\mu$&$\nu_{\mu}$ \\
& $t$&$b$&$\tau$&$\nu_{\tau}$\\
\hline
\hline new
fermions& $t^{'}$&$b^{'}$&$\tau^{'}$&$\nu'$ \\
\hline
\end{tabular}
\end{center}
\caption{The elementary particle spectrum of SM4}
\end{table}

If there exists a very heavy fourth neutrino $\nu'$,  it can
contribute to $a_\mu$ through diagram of Fig. 1. 
This is a
electroweak interaction. Similar to that of quarks, the
corresponding Lagragian is
\begin{equation}
{\cal L}=-\frac{g}{\sqrt{2}}({\bar \nu'}\gamma_\mu a_L V^l_{2\nu'} \mu) W^\mu
+h.c.
\end{equation}
where $a_L =(1-\gamma_5)/2$, $V^l_{2\nu'}$ is the (2,4) element of
the four-generation CKM matrix ($4\times 4$),
 \begin{equation}
V^{\rm SM4}_{\rm CKM} = \left (
\begin{array}{lcrr}
V_{1\nu_e} & V_{1\nu_\mu} & V_{1\nu_\tau} & V_{1\nu'}\\
V_{2\nu_e} & V_{2\nu_\mu} & V_{2\nu_\tau} & { V_{2\nu'}}\\
V_{3\nu_e} & V_{3\nu_\mu} & V_{3\nu_\tau} & V_{3\nu'}\\
V_{4\nu_e} & V_{4\nu_\mu} & V_{4\nu_\tau} & V_{4\nu'}\\
\end{array} \right )
\end{equation}

Reverting back to the diagrams of Fig. 1, we see that the fourth neutrino contribution to
$a_\mu$ is
\begin{eqnarray}
a^{\rm SM4}_\mu &=& \alpha(\frac{g}{\sqrt{2}}V^l_{2\nu'})^2
\frac{2m^2_\mu}{m^2_W}\cdot f(x)
 ,
\end{eqnarray}
where $x\equiv m^2_{\nu'} /m^2_W$, $\alpha$ is the fine
construction constant and
\begin{eqnarray}
 f(x)=\frac{-5x^3 -5x^2 +4x}{12(x-1)^3}+\frac{(2x^3 -x^2 )\log x}{2(x-1)^4}.
\end{eqnarray}

We suppose that the 1.6$\sigma$ discrepancy of muon anomalous
magnetic moment, $\delta a^{\rm SM}_\mu$, is induced by the fourth
sequential neutrino $\nu'$. We can  use the above equation to get
parameter space of $f(x)$ and $V^l_{2\nu'}$ to $m_{\nu'}$ (see
Fig. 2 and 3),
\begin{eqnarray}
(V^l_{2\nu'})^2 =\frac{{\sqrt{2}}\cdot \delta a^{\rm SM}_\mu}{ 8G_{\rm F}
\alpha m^2_\mu\cdot f(x)}
 \end{eqnarray}
From Fig. 2, we can see that  $f(x)$ is taken negative values
except for a very narrow range of $m_{\nu'}$ which is  from
58$GeV$ to 80$GeV$. In other words, the sign of $a^{\rm SM4}_\mu$
is only related to $\nu$ mass.  Only in this narrow mass range,
$\nu'$ gives positive contribution to $g_\mu -2$. The low bound of
$m_{\mu'}$ we get from $g_\mu -2$ is consistent with the present
experiments \cite{Mark}. But the upper bound, $m_{\mu'}< 80 {\rm
GeV}$, seems to conflict with the current experiments statue which
there is no any new physics signals upper to several GeVs. The
fourth generation particles seems not to be so light.  Moreover,
from Fig. 3,  if we consider the unitarity of the matrix
$V^{SM4}_{CKM}$ and tiny values of its elements, the reasonable
value range of $m_\mu'$ will be more narrow.

 In summary, we calculate the contribution of the fourth generation
to $g-2$ and get an interesting result which we can exclude most
values of $m_{\nu'}$.  Considering the new revised deviation of
$g-2$, we give the parameter space of $m_{\mu'}$ and lepton mixing
matrix element $V_{2\mu'}$. We find that $g-2$ can constrain on
the neutrino mass of the fourth generation: i.e. its mass should
be heavier than 58 GeV and lighter than 80 GeV. It seems that from
the lepton part, the current experiments can impose a stringent
constraint on the existence of the fourth generation.

\section*{Acknowledgments}
This research is supported  by the the Chinese Postdoctoral
Science Foundation and CAS K.C. Wong Postdoctoral Research Award  Fund.
 We are grateful to prof. C.S. Huang and Prof. X.M
Zhang for useful discussions.

\newpage
\begin{figure}
\vskip 14cm
\epsfxsize=20cm
\epsfysize=10cm
\centerline{
\epsffile{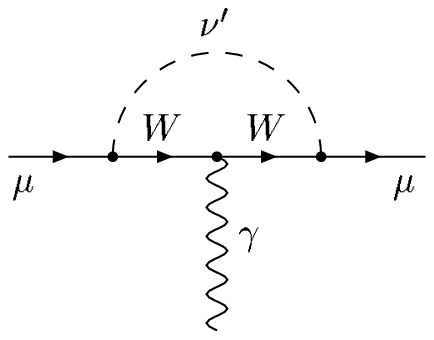}}
\vskip -18cm
\caption{ Feynmann diagram for $a_\mu$ induced by $\mu'$.}

\end{figure}
\newpage
\begin{figure}
\vskip 7cm
\epsfxsize=20cm
\epsfysize=10cm
\centerline{
\epsffile{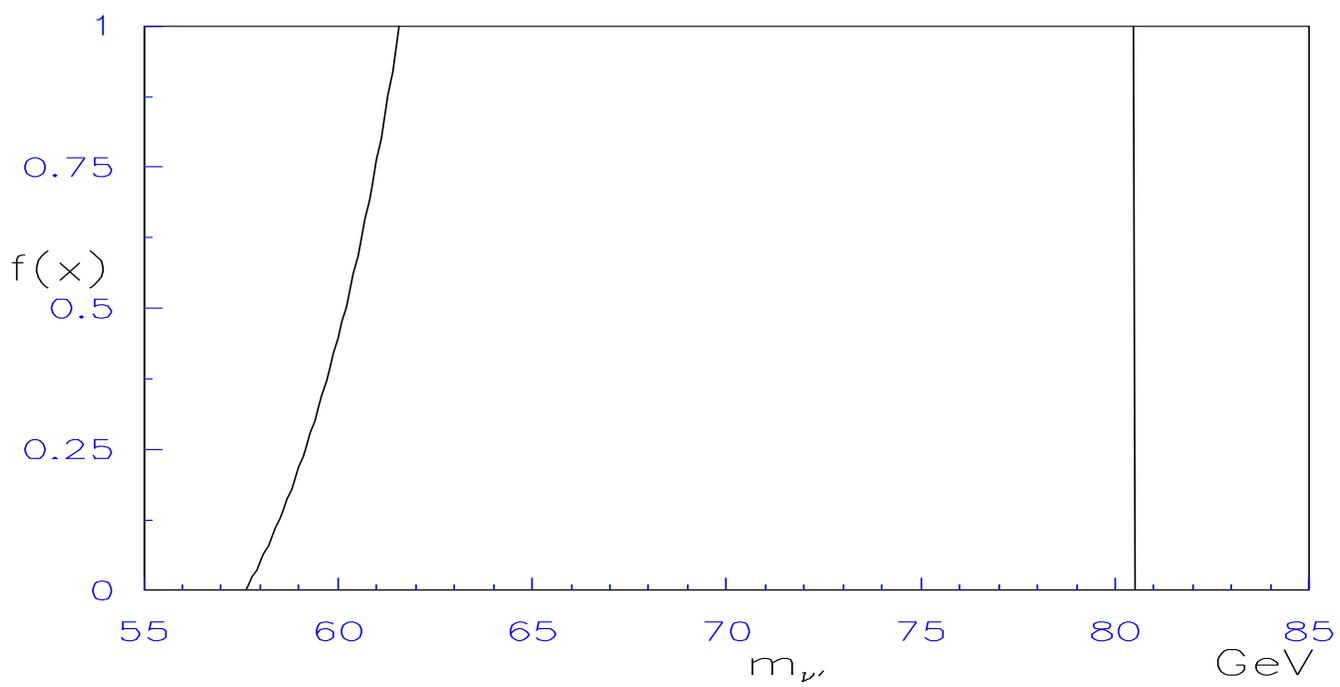}}
\caption{ Diagram of f(x) to $m_{\mu'}$ }
\end{figure}
\newpage
\begin{figure}
\vskip 7cm
\epsfxsize=20cm
\epsfysize=10cm
\centerline{
\epsffile{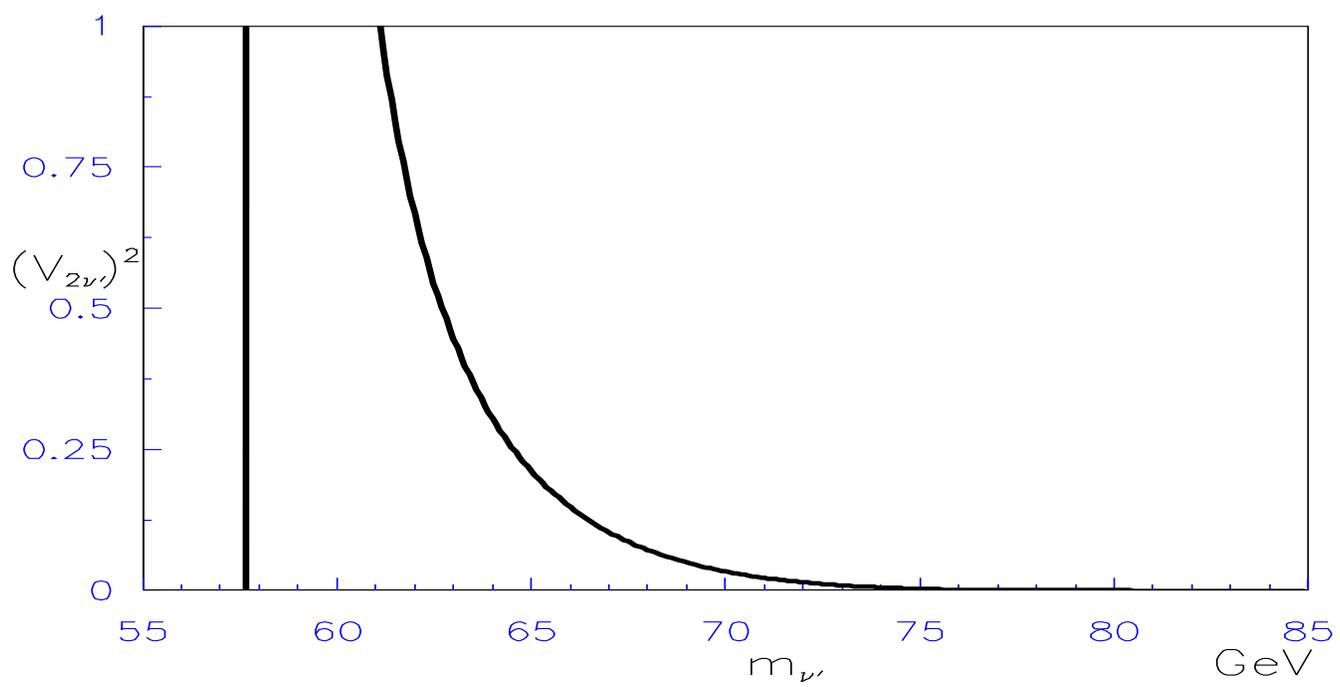}}
\caption{ Diagram of$V_{2\mu'}$ to $m_{\mu'}$ .}
\end{figure}

\end{document}